\documentclass[preprint]{revtex4}
\usepackage{amssymb}
\usepackage{amsmath}
\usepackage{graphicx}
\usepackage{epstopdf}
\begin{document}

\title{{\bf Hot magnetized nuclear matter: Thermodynamic and Saturation Properties}}

\author{{\bf  Z. Rezaei$^{1}$ \footnote{Corresponding author. E-mail:
zrezaei@shirazu.ac.ir}}, {\bf G. H. Bordbar$^{1,2}$}}
 \affiliation{ $^1$Department of Physics and Biruni Observatory,
Shiraz University,
Shiraz 71454, Iran\footnote{Permanent address},\\
 and\\
$^2$Research Institute for Astronomy and Astrophysics of Maragha,
P.O. Box 55134-441, Maragha 55177-36698, Iran}


\begin{abstract}
We have used a realistic nuclear potential, $AV_{18}$, and a many body technique, the lowest order
constraint variational  (LOCV) approach, to calculate the properties of hot magnetized
nuclear matter. By investigating the free energy, spin polarization parameter, and symmetry energy, we have studied the temperature and magnetic field dependence of the saturation properties of magnetized nuclear matter. In addition, we have calculated the equation of state of magnetized nuclear matter at different temperatures and magnetic fields. It was found that the flashing
temperature of nuclear matter decreases by increasing the magnetic field. In addition, we have studied the effect of the magnetic field on liquid gas phase transition of nuclear matter. The liquid gas coexistence curves, the order parameter of the liquid gas phase transition, and the properties of critical point at different magnetic fields
have been calculated.
\end{abstract}



\maketitle

\section{INTRODUCTION}

Dense nuclear matter at high temperatures and strong magnetic fields
can be found in the formation of protoneutron stars and relativistic heavy ion
collisions. In such systems, the temperature reaches $20-50 \ MeV$ \cite{Chapline,Camenzind}
as well as magnetic fields up to $10^{19}\ G$ \cite{Yuan98}. Therefore, investigation of the hot and magnetized nuclear matter is one of the important problems in both astrophysics and nuclear physics.

Many authors have studied the properties of nuclear
matter at high temperatures \cite{Barranco,Das,Kanzawa,Ghulam,Mukherjee,Ghodsi}.
Applying a finite temperature nuclear
Thomas-Fermi model with a Wigner-Seitz approximation,
the clustering of nucleons in hot dense matter has been studied \cite{Barranco}.
It has been concluded that the thermodynamic variables are fairly insensitive to the exact
arrangement of the nucleons, while the sizes of the
nuclear clusters exhibit an uncomfortable sensitivity
to relatively small effects.
A fully self-consistent model with an effective interaction has been used to derive the temperature dependence of different thermodynamic quantities
of nuclear matter \cite{Das}. It has been found that the calculated entropy is in agreement with the experiment.
The variational method and realistic nuclear potentials have also been applied to construct
the equation of state for nuclear matter at finite temperatures \cite{Kanzawa}.
Within the static fluctuation approximation and using the Reid68 and Reid93 soft-core
potentials, the bulk and thermodynamic properties of nonrelativistic hot nuclear matter
have been considered \cite{Ghulam}.
A variational theory for fermions at finite temperature and high density has been applied to study the equation of state of symmetric nuclear matter and the behavior of the nucleon effective mass in medium, and the fate of the neutral pion condensation at finite temperature \cite{Mukherjee}.
The equation of state of hot polarized nuclear matter has been applied to simulate the repulsive force caused by the incompressibility effects of nuclear matter in the fusion reactions of heavy colliding ions  \cite{Ghodsi}. The results show that the temperature effects of compound nuclei have significant importance in simulating the repulsive force on the fusion reactions.

One of the important quantity in nuclear matter to study
the ferromagnetic phase transition, the equation of state and the structure of
neutron rich nuclei and protoneutron stars is the symmetry energy of nuclear matter.
It has been concluded that there is a correlation of the crust-core transition density
and pressure in neutron stars with the slope of the symmetry energy and
the neutron skin thickness \cite{Pais}. In addition, the sensitivity of nuclear
symmetry energy elements at the saturation density to the binding
energies of ultra neutron-rich nuclei and the maximum mass of neutron
star has been also considered \cite{Mondal}. It has been shown that
a strong correlation of the neutron star radii with the
linear combination of the slopes of the nuclear matter incompressibility
and the symmetry energy
coefficients at the saturation density can be exist \cite{Alam}.
Since the nuclear matter can be at high temperatures, it is necessary to investigate
the symmetry energy of hot nuclear matter.
Applying a consistent Hartree-Fock study of the equation of state (EOS) of asymmetric nuclear matter
at finite temperature, it has been confirmed that there is
a strong
impact of the symmetry energy and nucleon effective mass on thermal properties and composition
of hot protoneutron star matter \cite{Tan}.
In the framework of the local density approximation, the temperature dependence of the symmetry energy for isotopic chains of even-even Ni, Sn, and
Pb nuclei has been explored \cite{Antonov}.The
results show that the symmetry energy decreases with temperature.

Studying the liquid gas phase transition of hot nuclear matter has also been considered
in some researches \cite{Natowitz,Elliott,Friedman,Malheiro,Baldo,Abd-Alla,Zuo3,Zuo,Rios2010,Sharma,Zhang,Rios8,Wellenhofer,Rios}.
According to the data from the heavy ion collisions  \cite{Natowitz} and nuclear reaction experiments \cite{Elliott}, the critical temperature of the liquid gas phase transition is $16.6\ MeV$ and  $17.9\  MeV$, respectively.
This is while the models and interactions affect the critical temperature obtained using the theoretical calculations. The results are $17.5\ MeV$  \cite{Friedman}, $13.6-18.3\ MeV$ \cite{Malheiro}, $21\ MeV$ and $20\  MeV$ \cite{Baldo}, $16.2\ MeV$ \cite{Abd-Alla}, $15-20\ MeV$ \cite{Zuo3}, $13\ MeV$ \cite{Zuo}, $14-23\ MeV$ \cite{Rios2010},  $14.7\ MeV$ \cite{Sharma}, and $15.7\ MeV$ \cite{Zhang} using different models.
Within the Self-Consistent Green's Functions approach and
employing $AV_{18}$ potential, the results for the
critical and flashing temperatures are
$11.6\ MeV$ and $9.5\ MeV$ \cite{Rios8}.
Applying microscopic nuclear forces from the
chiral effective field theory, the critical
temperature has been determined to be between $17.2\ MeV$ and
$19.1\ MeV$ \cite{Wellenhofer}.
The thermal properties and the liquid gas phase transition of symmetric nuclear matter have been also investigated using a standard covariance analysis \cite{Rios}. The results for the critical temperature lie  between $14 \ MeV$ and $18\ MeV$. Besides, the flashing temperature is within $11 \ MeV$ and $13\ MeV$.

Strong magnetic fields created in dense nuclear media also affect the thermodynamic properties of such systems. Different studies have explored the nuclear matter in the presence of strong magnetic fields \cite{Chakra7,Rabhi,Diener,Dong2,Haber,Aguirre}.
In a relativistic Hartree theory, it has been indicated that the application of magnetic field leads to additional binding for the system with a softer
equation of state \cite{Chakra7}. Besides, the saturation density
of nuclear matter increases by increasing the magnetic
field.
Using the relativistic nuclear models, it has been found that the presence
of the magnetic field will increase the instability region \cite{Rabhi}.
Effective baryon-meson exchange models with magnetic field coupled to the charge
and the dipole moment of the baryons have also confirmed that
by increasing the magnetic field, the saturation
density of nuclear matter increases and the system
becomes less bound \cite{Diener}. In addition, it has been shown
that as the magnetic field increases, the system becomes more incompressible.
In the framework of the relativistic mean field models FSU-Gold, it has been confirmed that
at low densities and by increasing the magnetic field, the energy per particle turns out to be increasing lower and a softening of the equation of state appears  \cite{Dong2}. However, at high densities, while the softening of the equation of state (EOS) will be gradually overwhelmed by stiffening resulting from the anomalous magnetic moments effect, the energies are slightly reduced by a strong magnetic field.
Relativistic field-theoretical models for nuclear matter have been applied to study the creation of nuclear matter in a sufficiently strong magnetic field \cite{Haber}. It has also been clarified that nuclear matter is more strongly bound in a magnetic field.
The equation of state, the compressibility, and magnetic susceptibility of nuclear matter in the presence of a magnetic field have been investigated employing the non-relativistic Skyrme potential model within a Hartree-Fock approach \cite{Aguirre}.

One of the many-body approaches in nuclear systems is the lowest order
constraint variational (LOCV) method \cite{Owen6,Owen74,Owen7}. This approach has been extended
to the finite temperature for neutron matter, nuclear
matter, and asymmetrical nuclear matter \cite{Modarres3,Modarres5,Modarres7,Moshfegh}. This method
is fully self-consistent, with no free parameters into the calculations.
In addition, this approach applies a normalization constraint \cite{Feenberg}
leading to small values for the higher-order terms \cite{Owen7,Bordbar97,Moshfegh}.
Besides, to perform an exact functional
minimization of the two-body energy with respect to the
short-range behavior of the correlation function, a particular form for the long-range behavior of
the correlation function is assumed. Therefore, we obtain a computational simplification
over the unconstrained methods to parameterize the short-range behavior of the correlation
functions.
In our previous studies, we have investigated the polarized nuclear matter at finite temperature applying LOCV method using realistic nuclear potentials in the absence of the magnetic field \cite{BB75,BB76,BB77,BB78,Bordbar80,Bordbar15}. In addition, we have recently studied the properties of magnetized nuclear matter at zero temperature \cite{Bordbar16,Bordbar413}.
In this paper, we are interested in the saturation and thermodynamic properties of magnetized nuclear matter at finite temperatures using LOCV method applying $AV_{18}$ potential.


\section{LOCV calculations for magnetized nuclear
matter at finite temperature} 

In this work, we propose a system of pure homogeneous symmetric nuclear
matter at finite temperature in the presence of a uniform magnetic field $B$
along the $z$-axis. Our system contains $A$ nucleons with spin-up (+) and spin-down (-).
The number densities of nucleons with the isospin and spin projection $j$ and $i$, respectively, are denoted by
 $\rho_j^{(i)}$. In addition, $\rho^{(i)}=\rho_p^{(i)}+\rho_n^{(i)}$ presents the number density of nucleons
with the spin projection $i$.  We introduce the spin
polarization parameter by $\delta=(\rho^{(+)}-\rho^{(-)})/ \rho$
in which $-1\leq\delta\leq1$, and $\rho=\rho^{(+)}+\rho^{(-)}$ is the
 total density of
system.

To study the macroscopic properties of the system, we
calculate
the total free energy per nucleon, ${\cal F}$,
\begin{eqnarray}\label{free}
      {\cal F}=E-T(S_n^{(+)} +S_p^{(+)}+S_n^{(-)}+S_p^{(-)}),
\end{eqnarray}
where $E$ is the total energy per nucleon and $S_j^{(i)}$ shows the entropy per
nucleon for the isospin and spin projection $j$ and $i$, respectively,
\begin{eqnarray}
 S_j^{(i)}(\rho,T,B)&=&-\frac{1}{A}\sum _{k}
 \left\{\left[1-n_j^{(i)}(k,T,
\rho_j^{(i)},B)\right]\times\right.
\nonumber \\&& \left.\ln \left[1-n_j^{(i)}(k,{T},\rho_j^{(i)},B)\right]
 +\right.
\nonumber \\&& \left.n_j^{(i)}(k,T,\rho_j^{(i)},B) \ln  n_j^{(i)}
(k,T,\rho_j^{(i)},B)\right\}.
\end{eqnarray}

In the above equation,  $n_j^{(i)}(k,T,\rho_j^{(i)},B)$ denotes the Fermi-Dirac
distribution function,
\begin{eqnarray}\label{fddf}
n_j^{(i)}(k,T,\rho_j^{(i)},B)&=&\frac{1}{\exp\left({\left[\epsilon_j^{(i)}
-\mu_j^{(i)}\right]/k_BT}\right)+1
},\end{eqnarray}
where $\epsilon_j^{(i)}(k,T,\rho_j^{(i)},B)$ and $\mu_j^{(i)}(T,\rho_j^{(i)},B)$ are the single particle
energy and chemical potential, respectively. For a system
at temperature $T$ and number density $\rho_j^{(i)}$,
the chemical potential is specified by
 \begin{eqnarray}\label{chpt}
 \sum _{k}
 n_j^{(i)}(k,T,\rho_j^{(i)},B)=N_j^{(i)}.
 \end{eqnarray}
Here, we approximate the single particle energy of nucleons with
the relations in terms of the effective mass and  the momentum independent
single particle potential,  $U_j^{(i)}(\rho,T,B)$, \cite{apv,dapv}.

The single particle energy of neutrons is
\begin{eqnarray}\label{spen}
 \epsilon_n^{(i)}(k,T,\rho_n^{(i)},B)&=&
 \frac{\hbar^{2}{k^2}}{2{m_n^{*}}^{(i)}
(\rho,T,B)}-\lambda_i \mu_n B+\nonumber \\&&U_n^{(i)}(\rho,T,B),
\end{eqnarray}
where $\lambda_\pm=\pm1$ and $\mu_n$ is the neutron magnetic moment. For the protons, we also have
\begin{eqnarray}\label{spep}
 \epsilon_p^{(i)}(k,T,\rho_p^{(i)},B)&=&
 \frac{\hbar^{2}{k^2}}{2{m_p^{*}}^{(i)}
(\rho,T,B)}+\nonumber \\&& \frac{e \hbar^{}{B}}{c{m_p^{*}}^{(i)}
(\rho,T,B)}(l^{(i)}+\frac{1}{2})-\lambda_i \mu_p B+\nonumber \\&&U_p^{(i)}(\rho,T,B).
\end{eqnarray}
Here, $l^{(i)}=0,\ 1,\ 2,\ 3,\ ...$ are the integers labeling the Landau levels \cite{Pathria} for a
proton with spin projection $i$, $e$ is the proton charge,  $c$ is the speed of light, and $\mu_p$ is the proton magnetic moment.

In our calculations, we numerically minimize the free energy with respect to the variations in
the effective masses, and we obtain the chemical potentials and the effective
masses of the spin-up and spin-down nucleons at the minimum point of the free energy.

The total energy of magnetized nuclear matter is calculated using the LOCV method.
In this approach, we consider a trial many body wave function of the form
\begin{eqnarray}
     \psi=F\phi,
 \end{eqnarray}
in which $\phi$ is the uncorrelated ground-state wave function of $A$
independent nucleons and $F$ is a proper $A$-body correlation
function.
Applying Jastrow ansatz \cite{Jastrow}, $F$ is
replaced by
\begin{eqnarray}
    F=S\prod _{i>j}f(ij),
 \end{eqnarray}
with the symmetrizing operator $S$.
The cluster expansion of the
 energy functional up to the two-body term
is as follows,
 \begin{eqnarray}\label{tener}
           E([f])=\frac{1}{A}\frac{\langle\psi|H|\psi\rangle}
           {\langle\psi|\psi\rangle}=E^p _{1}+E^n _{1}+E _{2}.
 \end{eqnarray}
$E^p _{1}$ and $E^n _{1}$ are the one-body energies of protons and neutrons, respectively,
and $E _{2}$  is the two-body energy.

At the temperature $T$, the one-body term for
the protons, $E^p _{1}$, is given by
\begin{eqnarray}\label{E1p}
 E^p _{1}&=&\frac{e B}{\pi h c \rho}\sum_{i=+,-} \int_0^{\infty} dk \sum_{l^{(i)}=0}^{\infty} n_p^{(i)}(k,T,\rho_p^{(i)},B)\nonumber \\&&\times ( \frac{\hbar^{2}{k^2}}{2{m_p^{}}^{}}+ \frac{e \hbar^{}{B}}{c{m_p^{}}^{}}(l^{(i)}+\frac{1}{2})-\lambda_i \mu_p B) ,
 \end{eqnarray}
where $n_p^{(i)}$  has been given in Eq. (\ref{fddf}). Moreover, the one-body term for
the neutrons, $E^n _{1}$, is as follows,
 \begin{eqnarray}
              E^n _{1}&=&\sum_{i=+,-}\sum _{k}
               n_n^{(i)}(k,T,\rho_n^{(i)},B) (\frac{\hbar^{2}{k^2}}{2m_n}-\lambda_i \mu_n B),
                \end{eqnarray}
with $n_n^{(i)}$  given in Eq. (\ref{fddf}).

The two-body energy, $E_{2}$, is
\begin{eqnarray}
    E_{2}&=&\frac{1}{2A}\sum_{ij} \langle ij\left| \nu(12)\right|
    ij-ji\rangle,
\end{eqnarray}
 where
\begin{eqnarray}
 \nu(12)&=&-\frac{\hbar^{2}}{2m}[f(12),[\nabla
_{12}^{2},f(12)]]+\nonumber \\&&f(12)V(12)f(12).
\end{eqnarray}
Here,  $V(12)$ denotes the nuclear potential which in this study, the $AV_{18}$ potential is substituted \cite{Wiringa}.
In addition, $f(12)$ presents the two-body correlation function with the form $f(12)=\sum^3_{k=1}f^{(k)}(r_{12})P^{(k)}_{12}$ with $P^{(k)}_{12}$ given in
Ref.~\cite{Bordbar80}.  The two-body energy can be obtained using the two-body correlation function and
the nuclear potential. Afterwards, the two-body energy is minimized with respect to the
variations in the functions $f^{(i)}$ subject to the
normalization constraint, $\frac{1}{A}\sum_{ij}\langle ij| h_{S_{z}}^{2}
-f^{2}(12)| ij\rangle _{a}=0$, with the Pauli
function $h_{S_{z}}(r)$ as follows \cite{Bordbar80},
\begin{eqnarray}
h_{S_{z}}(r)=
\left\{%
\begin{array}{ll}
\left[ 1-\frac{1}{2}\left( \frac{\gamma^{(i)}(r)
       }{\rho}\right) ^{2}\right] ^{-1/2} & ;\ \hbox{$S_{z}=\pm 1$} \\
    1 & ;\ \hbox{$S_{z}= 0$}
\end{array}%
\right.
\end{eqnarray}
 where
\begin{eqnarray}
\gamma^{(i)}(r)=\frac{1}{\pi^{2}}\int n^{(i)}(k,T,\rho^{(i)},B)J_{0}(kr)k^2dk .
 \end{eqnarray}
Solving the differential equations resulted from the minimization procedure leads to the correlation functions and the two-body energy term. In the next step, it is possible to calculate the free energy of nuclear matter using Eqs. (\ref{free}) and (\ref{tener}).
\section{RESULTS and DISCUSSION}\label{NLmatchingFFtex}

\begin{figure*}
\vspace*{5cm}       
\includegraphics{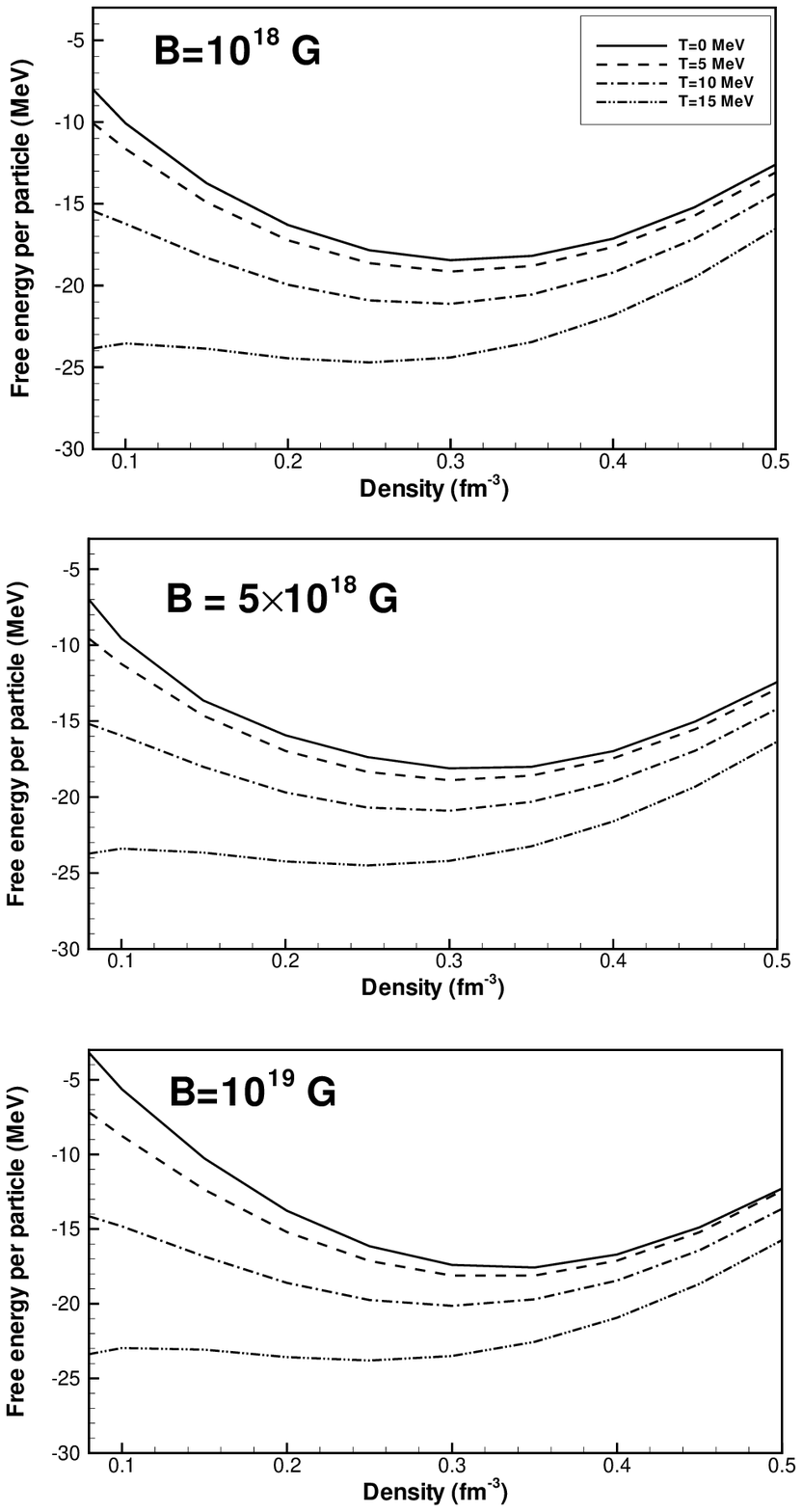}
\caption{{Free energy per particle versus the density at different temperatures and magnetic fields.}}
\label{fig1}       
\end{figure*}

\begin{figure*}
\vspace*{5cm}       
\includegraphics{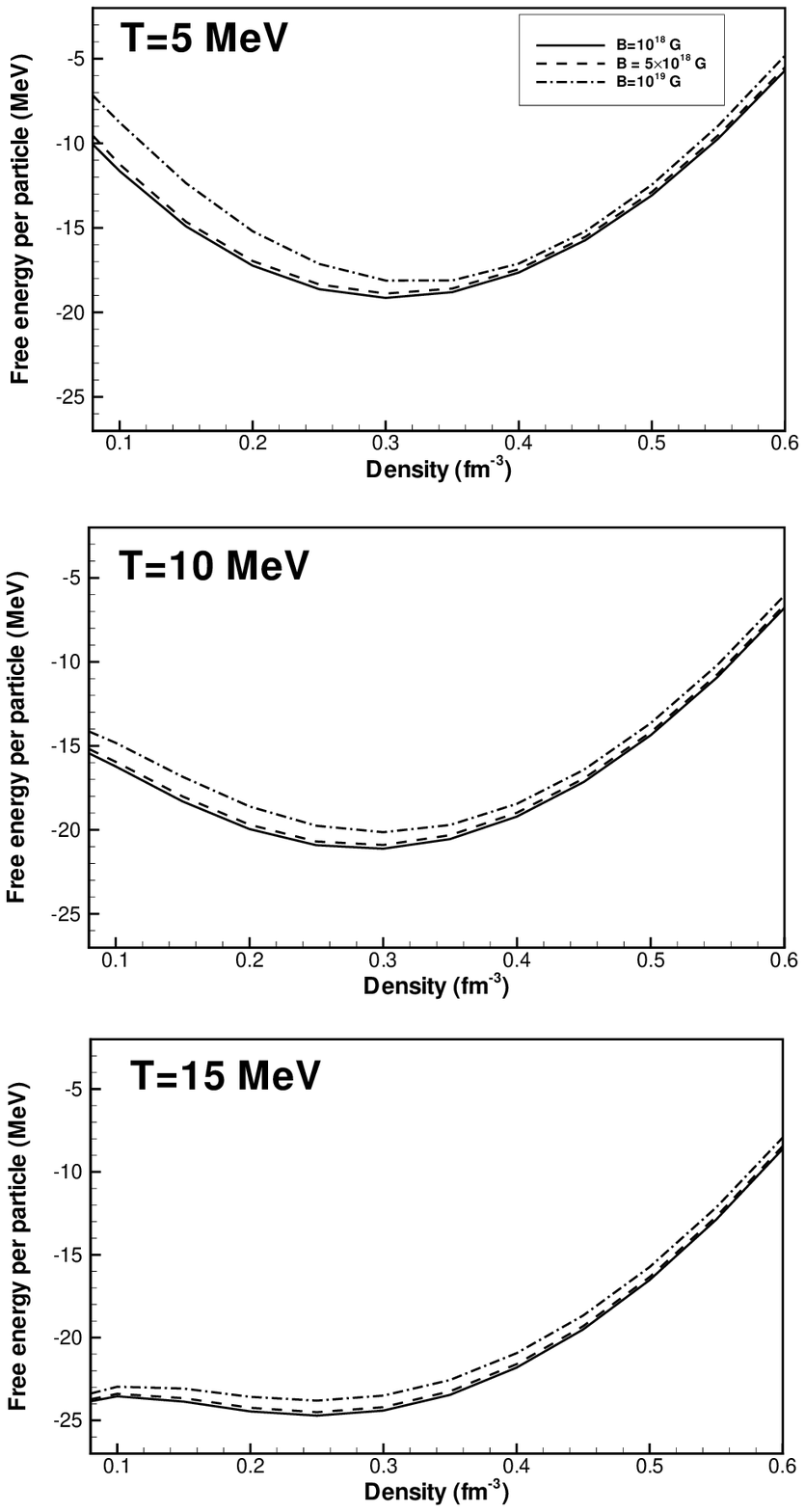}
\caption{{Same as Fig. \ref{fig1} but for the comparison of free energies at different magnetic fields.}}
\label{fig1B}       
\end{figure*}

\begin{figure*}
\vspace*{5cm}       
\includegraphics{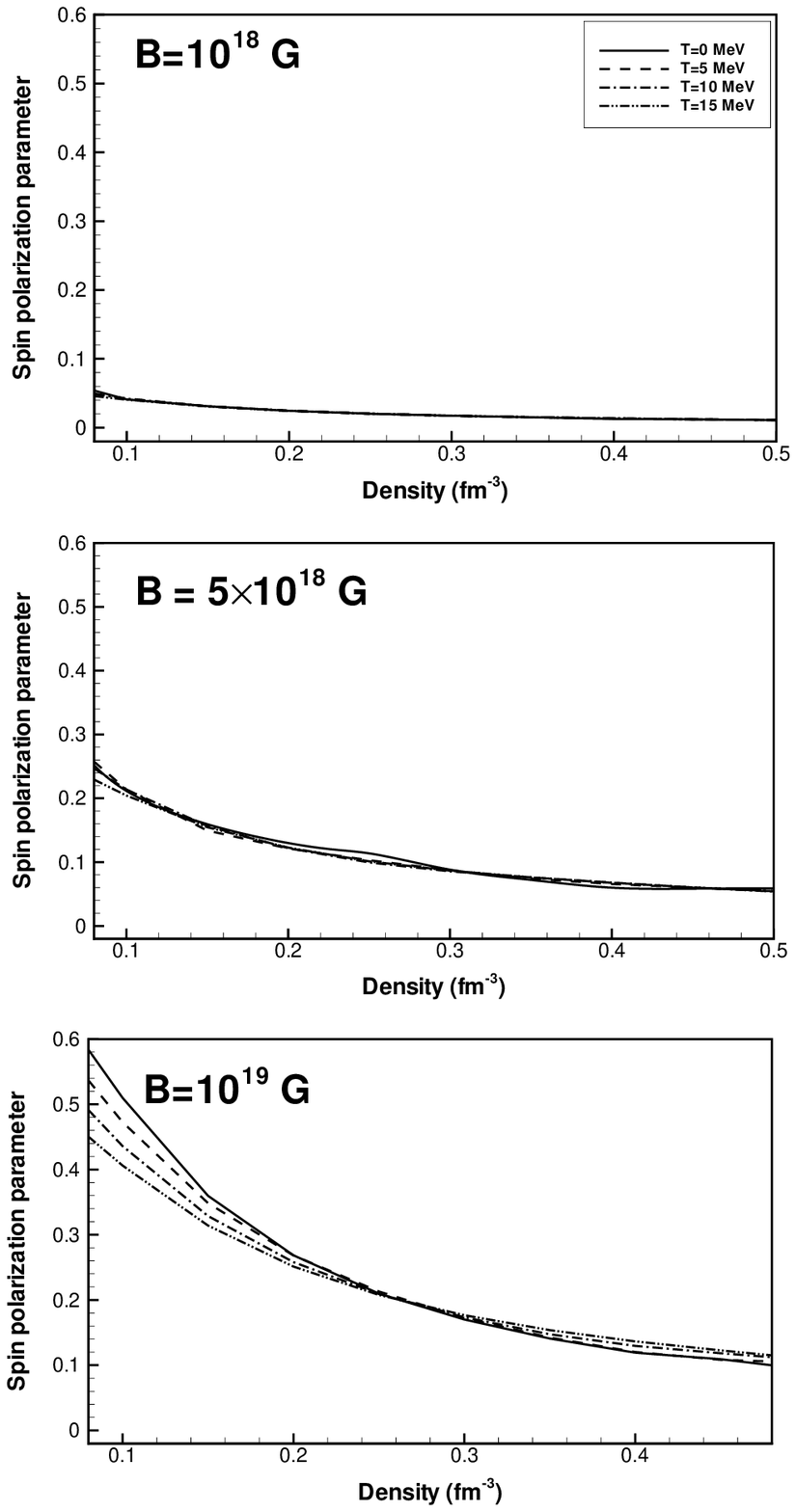}
\caption{Same as Fig. \ref{fig1} but for the spin polarization parameter.}
\label{fig2}       
\end{figure*}

To investigate the properties of hot magnetized nuclear matter, we have presented
the free energy of system as a function of the density at different temperatures and
magnetic fields in Figs. \ref{fig1} and \ref{fig1B}. It is obvious that for all given values of the magnetic field and temperature the nuclear matter
is self-bound. However, the density at which the nuclear matter saturates and the corresponding free energy depend on the temperature and magnetic field (as we will discuss these dependencies in more details). As we can see from Figs. \ref{fig1} and \ref{fig1B}, the influences of the temperature and magnetic field on the free energy are more considerable at lower densities. This is due to the larger contribution of nuclear energy at higher densities.  In addition, Fig. \ref{fig1B} indicates that the magnetic field can affect the nuclear matter more significantly at lower temperatures.
Fig. \ref{fig2} shows the density and temperature dependence of the nuclear matter spin polarization parameter
for different magnetic fields.
It is clear that the spin polarization parameter of the system decreases as density grows.
In addition, the spin polarization parameter of the nuclear matter has smaller values at higher temperatures, while it increases by increasing the magnetic field.
The effects of temperature and magnetic field are more significant at lower densities.

\subsection{Symmetry energy}

\begin{figure*}
\vspace*{5cm}       
\includegraphics{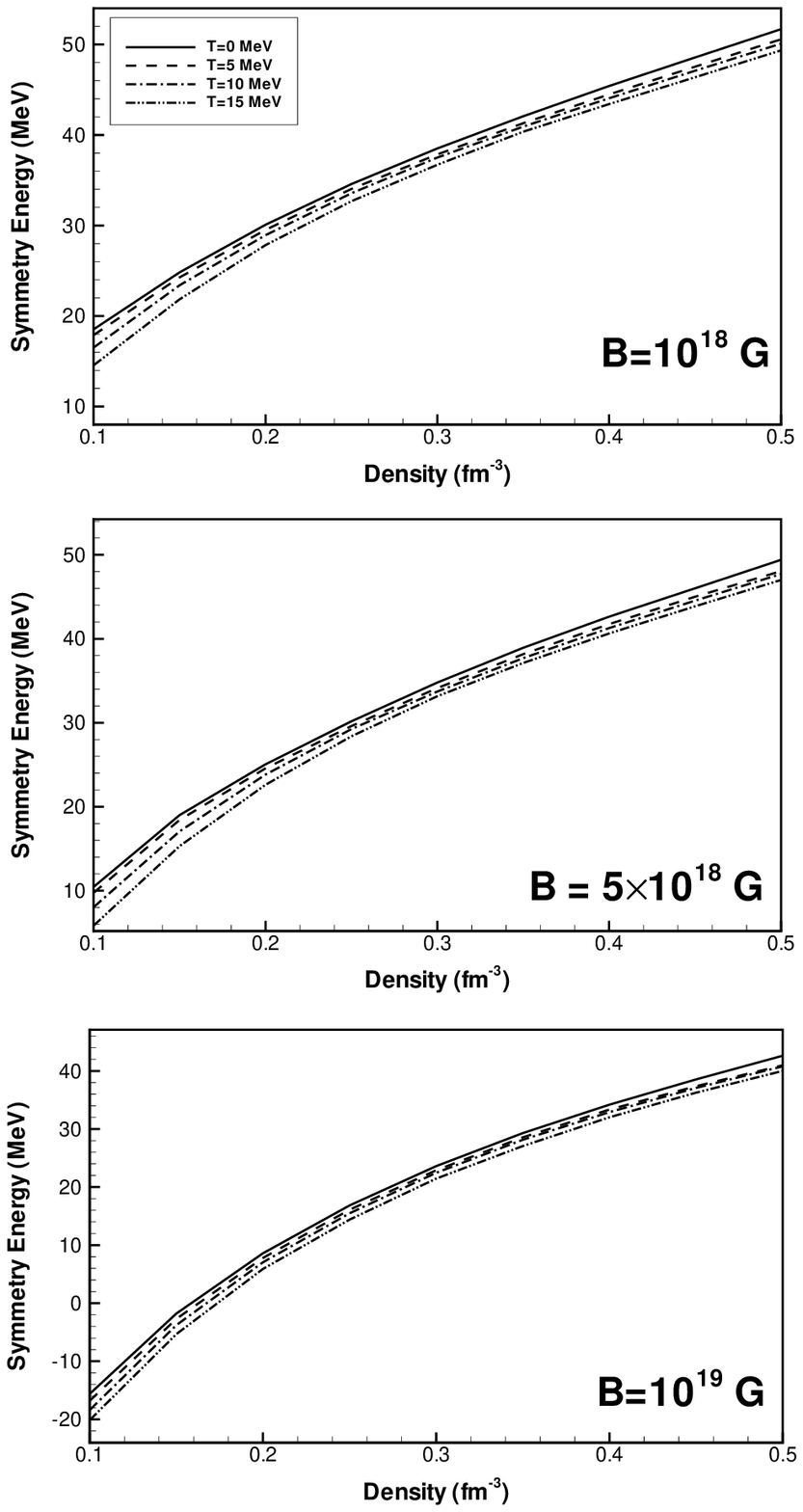}
\caption{{Same as Fig. \ref{fig1} but for the symmetry energy.}}
\label{figSE}       
\end{figure*}

\begin{figure*}
\vspace*{5cm}       
\includegraphics{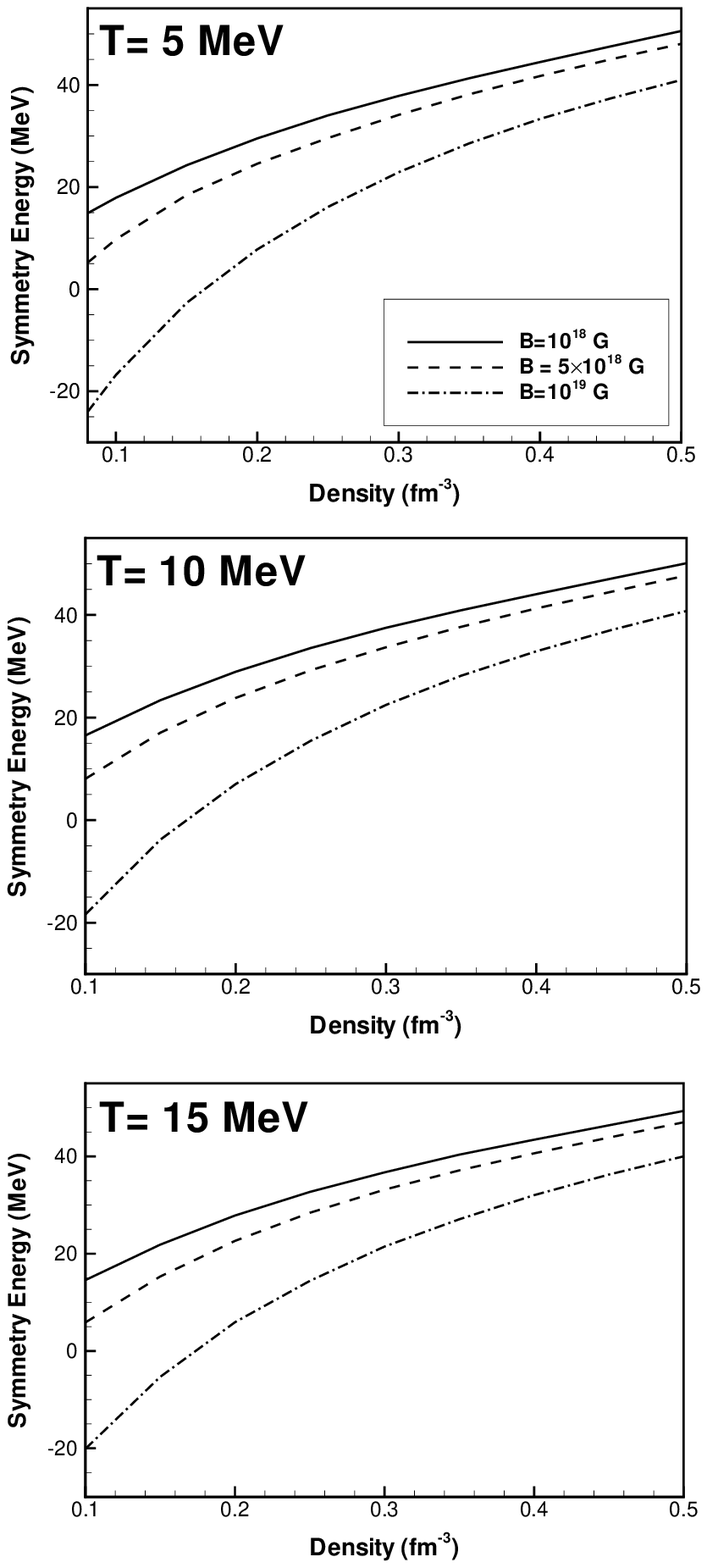}
\caption{{Same as Fig. \ref{fig1B} but for the symmetry energy.}}
\label{figSEb}       
\end{figure*}

For nuclear matter, the symmetry energy which quantifies the produced energy
when neutron matter converts to symmetric nuclear matter is
\begin{eqnarray}
    E_{sym}=\frac{1}{2}\frac{\partial^2E}{\partial\eta^2}\mid_{\eta=0},
\end{eqnarray}
where $\eta=(N-Z)/(N+Z)$ is the isospin asymmetry parameter.
In the parabolic approximation, the symmetry energy is given by \cite{Bordbar98}
\begin{eqnarray}
    E_{sym}=E_{neut}-E_{nucl}.
\end{eqnarray}
In the above equation, $E_{neut}$ and $E_{nucl}$ denote the energy per particle of the pure neutron matter and symmetric nuclear matter. Figs. \ref{figSE} and \ref{figSEb} present the symmetry energy of hot magnetized nuclear matter at different temperatures and magnetic fields. It can be seen that the symmetry energy grows as density increases. In addition, it is obvious from Fig. \ref{figSE} that at each magnetic field and density, the symmetry energy decreases as the temperature increases. This result is in agreement with the results of Ref. \cite{Antonov}. Fig. \ref{figSEb} confirms that the symmetry energy decreases with the increase in magnetic field. This quantity is more affected by the magnetic field at lower densities. Figs. \ref{figSE} and \ref{figSEb} show that the slope of the symmetry energy is larger at higher magnetic fields, while the temperature has a negligible effect on the slope of symmetry energy. Moreover, the results show that at high magnetic fields and low densities, the symmetry energy is negative. This is due to the interaction of nucleons magnetic moment and magnetic field.

\subsection{Saturation properties }
\begin{figure*}
\vspace*{5cm}       
\includegraphics{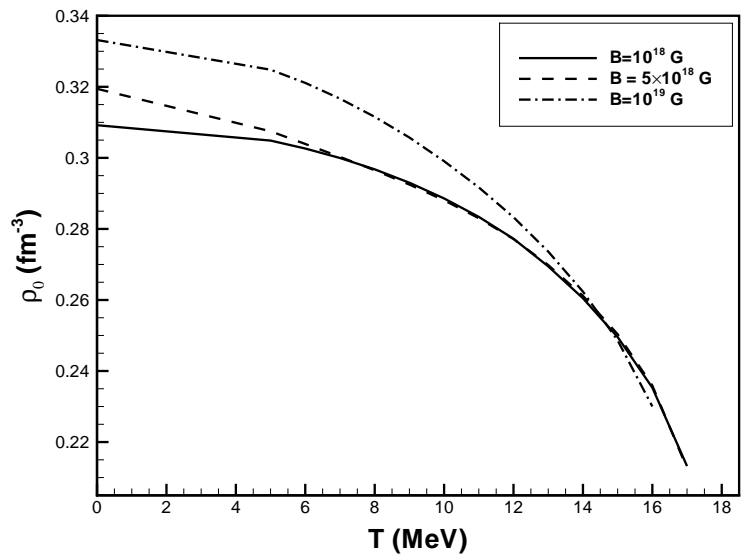}
\caption{{Saturation density as a function of temperature at different magnetic fields.}}
\label{fig3}       
\end{figure*}

We have plotted the saturation density of magnetized nuclear matter as a
function of temperature for different magnetic fields in Fig. \ref{fig3}. It is seen
that at each magnetic field, the saturation density decreases by increasing the
temperature. It means that the nuclear matter at higher temperatures
saturates at lower values of the density.
%
However, we can see that at low temperatures
for $B>10^{18}\ G$, the saturation density increases as the magnetic field grows. This result, for the nuclear matter at zero temperature, has been also reported in our previous study of cold magnetized nuclear matter \cite{Bordbar16}.
This increase is due to the fact that in the cold magnetized nuclear matter, the contribution of magnetic energy is significant and it prevents the nuclear matter from the saturation.  Therefore, in such conditions, the higher values of density are needed to saturate the nuclear matter. This is while for $T>15\ MeV$, the saturation density decreases by
increasing the magnetic field. This result indicates that in the nuclear matter at enough high temperatures, the interaction of nucleons with the magnetic field helps the system to saturate at smaller densities.

\begin{figure*}
\vspace*{5cm}       
\includegraphics{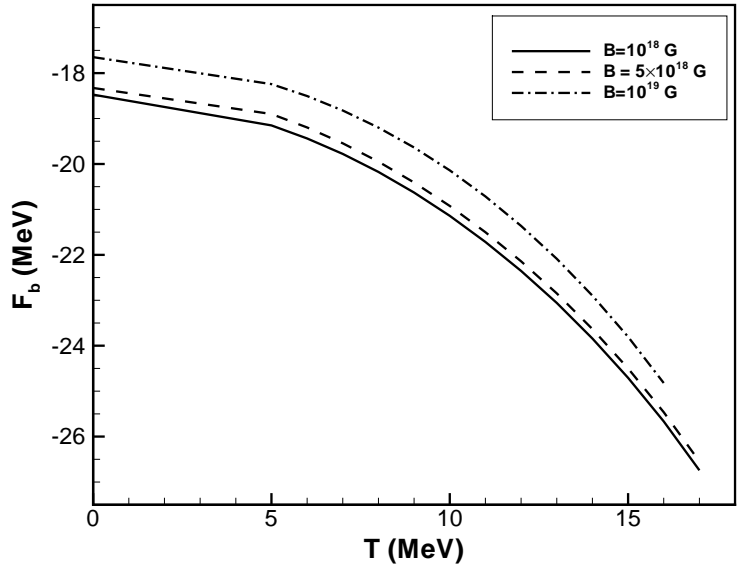}
\caption{{Temperature dependence of free energy per particle corresponding to the saturation point at different values of magnetic field.}}
\label{fig4}       
\end{figure*}

\begin{figure*}
\vspace*{5cm}       
\includegraphics{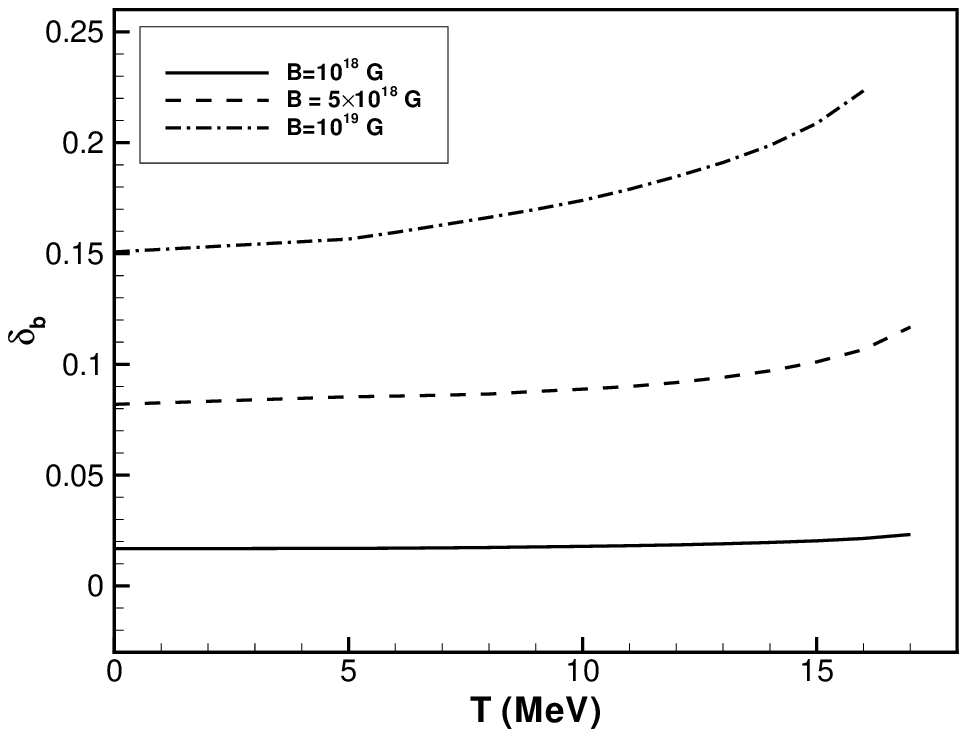}
\caption{{Same as Fig. \ref{fig4} but for the spin polarization parameter. }}
\label{fig5}       
\end{figure*}

Fig. \ref{fig4} presents the value of free energy at the saturation point versus the temperature
at different magnetic fields. It is obvious that this value of free energy
decreases by increasing the temperature. At each temperature, for $ B>10^{18}\ G$, the
free energy at the saturation point increases with the increase in magnetic field. This result also holds for cold nuclear matter \cite{Bordbar16}. This can be interpreted by the significant contribution of the magnetic energy at stronger magnetic fields.
Another result of this work is the behavior of the nuclear matter spin polarization parameter
at the saturation point which has presented in Fig. \ref{fig5}. At each
magnetic field, by increasing the temperature, the spin polarization parameter of system grows. This is
a consequence of the smaller values of saturation density at higher temperatures (Fig. \ref{fig3}) and the larger values of the nuclear matter spin polarization parameter at lower temperatures
(Fig. \ref{fig2}). In addition, the spin polarization parameter of nuclear matter increases by increasing the magnetic field.

\begin{figure*}
\vspace*{5cm}       
\includegraphics{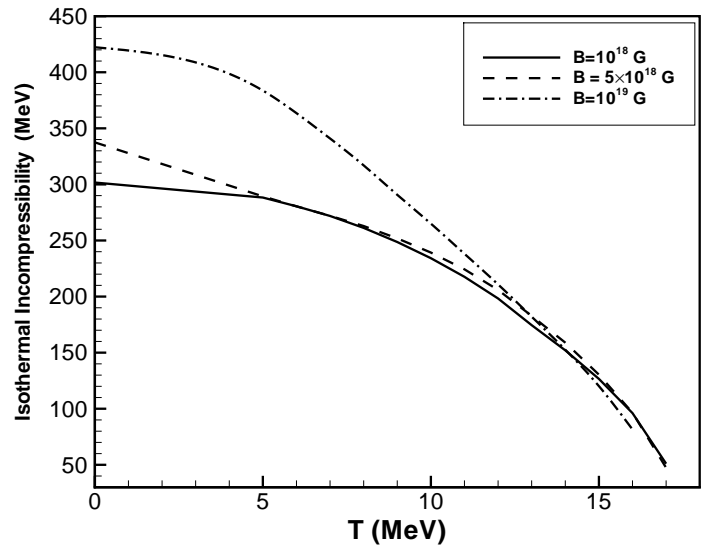}
\caption{{Same as Fig. \ref{fig4} but for the isothermal incompressibility. }}
\label{fig6}       
\end{figure*}

The isothermal incompressibility at the saturation point is given by
\begin{eqnarray}
\mathcal{K}(B,T )=9 \rho_0^2(B,T) [\frac{\partial^2{\cal F}(\rho,B,T)}{\partial \rho^2}]_{\rho=\rho_0(B,T)}.
 \end{eqnarray}
Fig. \ref{fig6} shows our results for the isothermal incompressibility of magnetized nuclear matter. At each magnetic field, the isothermal incompressibility
reduces by increasing the temperature. Previously, we have reported this result for zero magnetic field in Ref. \cite{Bordbar58}. It can be found from this result that as the temperature increases, the number of accessible states for the nucleons grows leading to
more compressible nuclear matter. However, it is clear from Fig. \ref{fig6} that at low temperatures and for $B>10^{18}\ G$,
the isothermal incompressibility increases by increasing the magnetic field.
This indicates the stiffening of equation of state (EOS) at high magnetic fields.
In fact, at low temperatures for $B>10^{18}\ G$, the softening of
EOS is overwhelmed by stiffening due to the spin polarization parameter
of nuclear matter, as in our previous study of cold magnetized nuclear matter \cite{Bordbar16}.
However, at high temperatures, the isothermal incompressibility decreases as the
magnetic field increases. This confirms the softening of EOS at high magnetic fields.
We can conclude that at high temperatures and magnetic fields, the Landau quantization is the dominant effect which leads to the softening of EOS.

\subsection{Equation of state }

\begin{figure*}
\vspace*{5cm}       
\includegraphics{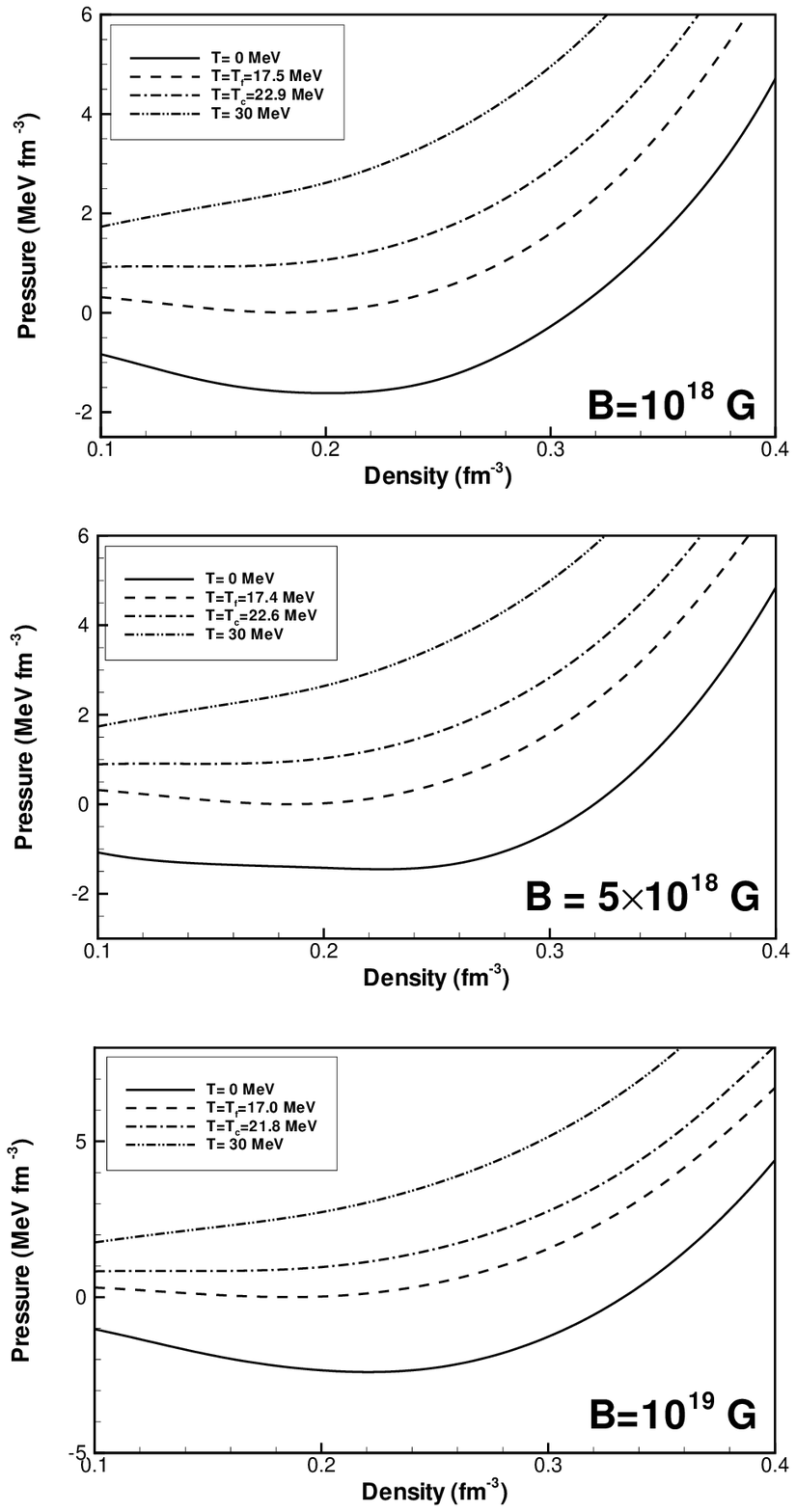}
\caption{{The equation of state of magnetized nuclear matter at different temperatures and magnetic fields.}}
\label{fig7}       
\end{figure*}

The equation of state for magnetized nuclear matter at different temperatures
and magnetic fields can be calculated using the
following relation,
\begin{eqnarray}
    P(\rho,T,B)= \rho^{2}
    {\left(\frac{\partial F(\rho,T,B)}
     {\partial \rho} \right)_{T,B}}.
\end{eqnarray}
Our results at four temperatures and three magnetic fields have
presented in Fig. \ref{fig7}. It is obvious that at zero temperature,
the pressure is negative below the saturation density. Besides, this quantity is zero at the saturation point.
For all magnetic fields, the range of density
at which the pressure is negative reduces as the temperature grows. This holds until
the pressure becomes a positive definite function at a specific point.
The temperature corresponding to this point is called flashing temperature,
$T_f$ \cite{Rios}. In fact, above the flashing temperature, the kinetic energy is more significant than the binding energy of the nuclear matter, and the increase of temperature makes the nuclear matter unbound.
Besides, the flashing density, $\rho_f$, which satisfies
the constraints,
\begin{eqnarray}
P(\rho=\rho_f)=(\frac{\partial P}{\partial \rho})_{\rho_f}=0,
 \end{eqnarray}
characterizes the flashing point.
The physical importance of the flashing point is due to the fact that
it corresponds to the maximum temperature of the nuclear matter at which
the system is self-bound \cite{Rios}. The nuclear matter at the flashing point
is at zero pressure. Because of the positive pressure of nuclear matter
at higher temperatures, \textit{i.e.} $T>T_f$, the increasing of temperature
from the flashing temperature leads to the expansion of nuclear matter.
The properties of the flashing point at different magnetic fields
are presented in Table~\ref{table1}. We can see that the flashing temperature,
with a value between $17.0$ and $17.5\ MeV$ for the given magnetic fields,
decreases as the magnetic field grows.
It is due to the fact that besides the temperature, the increase of the magnetic field
makes the system less bound. Therefore, for magnetized nuclear matter
at high magnetic fields, the low temperatures are needed to make the
system unbound. Our result for the
flashing density, \textit{i.e.} $0.19\ fm^{-3}$, indicates that this quantity is not sensitive to
the magnitude of magnetic field (see Table~\ref{table1}). Moreover,
our flashing density is greater than the half of our saturation density at zero
temperature and magnetic field, \textit{i.e.} $0.310\ fm^{-3}$ \cite{Bordbar98} which is similar to that of others \cite{Rios2010,Rios}.

By increasing the temperature above the flashing point, the isotherms are positive at each density.
However, at some densities, these isotherms are decreasing function of the density, \textit{i.e.} $\partial P/\partial \rho<0$.
This holds until we reach the critical temperature, $T_c$. For the critical temperature and the temperatures above that,
the pressure grows monotonically as the density increases.
We have presented the critical isotherm in Fig. \ref{fig7}.
The value of the critical density, $\rho_c$, can be found by the following constraints,
\begin{eqnarray}
(\frac{\partial P}
{\partial \rho})_{\rho_c}=(\frac{\partial^2 P}{\partial \rho^2})_{\rho_c}=0.
 \end{eqnarray}
The critical properties of magnetized nuclear matter will be discussed in the following. Fig. \ref{fig7} also shows an isotherm above the critical point, \textit{i.e.} $T = 30\ MeV$.

\subsection{Liquid gas phase transition}

\begin{figure*}
\vspace*{5cm}       
\includegraphics{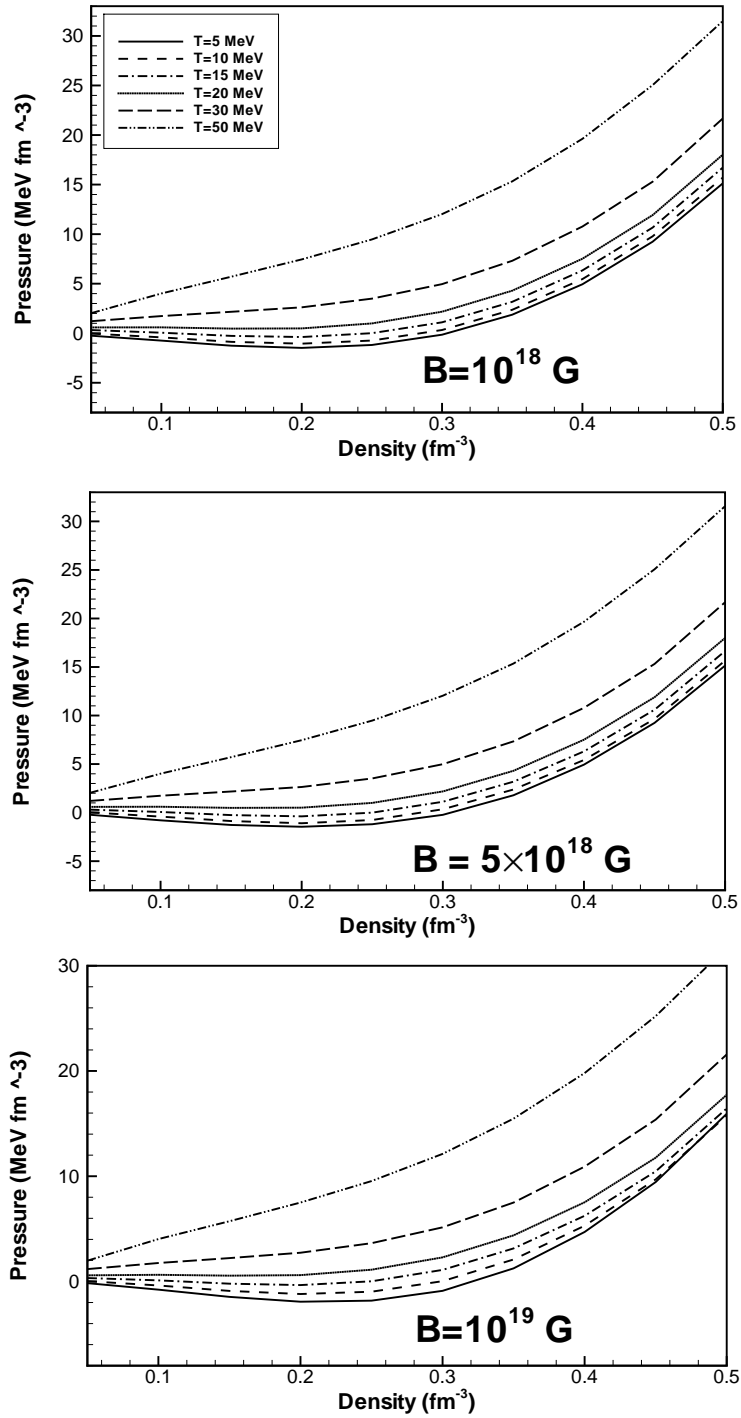}
\caption{{Same as Fig. \ref{fig7} but at other values of the temperature.}}
\label{fig8}       
\end{figure*}

\begin{figure*}
\vspace*{5cm}       
\includegraphics{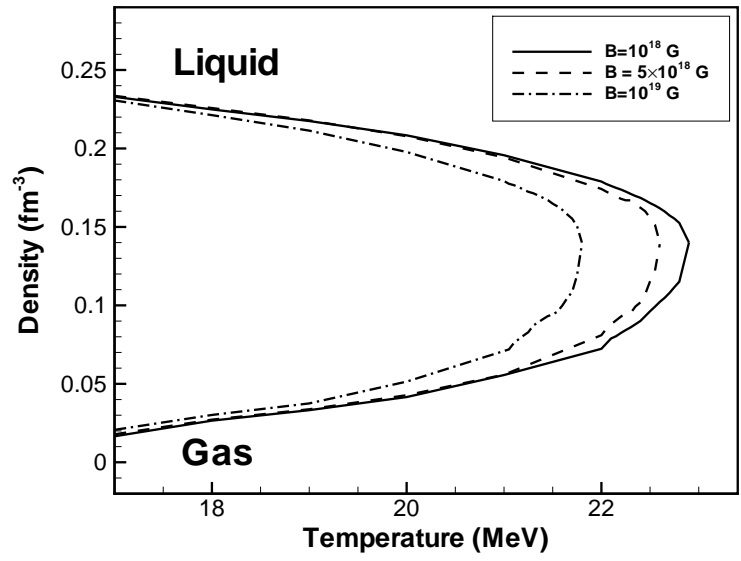}
\caption{{The liquid gas coexistence curve at different magnetic fields.}}
\label{fig9}       
\end{figure*}

Fig. \ref{fig8} gives the equation of state at different temperatures and magnetic fields.
We can find from this figure that at low temperatures and densities, the nuclear matter is mechanically unstable. This leads to a first order liquid gas phase transition in the nuclear matter.
It is clear that the isotherms present a typical Van der Waals like behavior in which the liquid and gaseous
phases coexist.
It is possible to find the densities of gas and liquid at each temperature
applying the equal-area Maxwell construction.
Fig. \ref{fig9} shows the coexistence curves, \textit{i.e.} the liquid and gas densities versus temperature at different magnetic fields. It is clear that the liquid density decreases and the gas density increases by increasing the temperature.
Inside the coexistence curve, the stable phase of nuclear matter is a mixture of liquid
and gas. The coexistence region reduces as the temperature increases, and
it disappears at the critical temperature. At this point, the densities of gas and liquid
become equal and the nuclear matter experiences a second-order phase transition.
It is clear that the coexistence region reduces as the magnetic field grows.
In addition, at high magnetic fields, the coexistence region disappears at lower temperatures. This shows that
the critical temperature reduces by increasing the magnetic field. Fig. \ref{fig9} also indicates that
the effect of magnetic field on the coexistence curve is more significant at higher temperatures.
The critical properties of magnetized nuclear matter
at different magnetic fields have been given in Table~\ref{table2}. We can see that by increasing the magnetic field from $10^{18}$ to $10^{19}\ G$,
the critical temperature decreases from $22.9$ to $21.8\ MeV$.
In addition,  the critical pressure decreases from $0.93\ MeV fm^{-3}$ to $0.84\ MeV fm^{-3}$ by increasing the magnetic field from $10^{18}\ G$ to $10^{19}\ G$. However, the critical densities are nearly identical for different magnetic fields.
According to the results given in Table~\ref{table2}, it can be easily seen that the value of the ratio $\gamma_c=P_c/(\rho_c T_c)$ is $0.29$, $0.28$, and $0.28$ for the magnetic fields $10^{18}$, $5\times10^{18}$, and $10^{19}\ G$, respectively, which are close to the value $\gamma_c\approx0.28$ \cite{Rios2010,Rios}. Besides, Table~\ref{table2} shows that as the magnetic field increases, the critical temperature and pressure decrease toward the experimental values $T_c=17.9\ MeV$ and $P_c=0.31\ MeV fm^{-3}$.

\begin{figure*}
\vspace*{5cm}       
\includegraphics{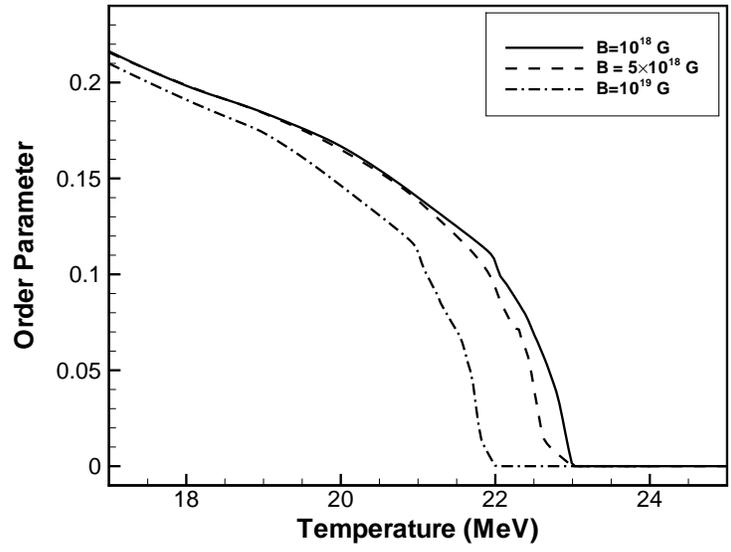}
\caption{{The order parameter for the liquid gas phase transition as a function of
temperature at different magnetic fields.}}
\label{fig10}       
\end{figure*}

The difference in the liquid and gas
densities denotes the order parameter of the liquid gas phase transition.
We define the order parameter as $m =
\rho_{liquid} -\rho_{gas}$ to show the behavior of liquid gas phase transition in the nuclear matter.
Fig. \ref{fig10} presents the order parameter versus the temperature at different magnetic fields.
It can be seen that for each magnetic field, the order parameter vanishes at the critical temperature.
Below the critical point, the order parameter decreases as the magnetic field grows.
The effects of the magnetic field on the order parameter is more important at higher temperatures.

\section{Summary and Concluding Remarks}

The microscopic nuclear potential, $AV_{18}$, and LOCV method were applied to investigate the properties of symmetric nuclear matter in the presence of strong magnetic fields at finite temperature, the conditions which are found in protoneutron stars and heavy ion collisions.
The free energy of nuclear matter versus the density at different temperatures
and magnetic fields shows that the nuclear matter is self-bound at low enough temperatures.
We have shown that the effect of the temperature and magnetic field on the free energy is
more considerable at lower densities. Moreover, due to the contribution of strong interaction at high densities, the effect of the magnetic field on the nuclear matter is more significant at lower temperatures. It has been shown that the symmetry energy decreases by increasing the temperature and magnetic field. Our results indicate that the saturation density decreases by increasing the temperature.
However, the effect of the magnetic
field on the saturation density depends on the temperature. At lower temperatures, the magnetic field increases the saturation density, while it decreases the saturation density
for the system at higher temperatures.
In addition, we found that the free energy corresponding to
the saturation point decreases as the temperature grows. Besides, the free energy at the saturation point increases with the increase in magnetic field.
The spin polarization parameter of nuclear matter at the saturation density is an increasing function of the temperature. This enchantment is the result of the lower values of the saturation density at higher temperatures.
It was clarified that the isothermal incompressibility is a decreasing function of the temperature.
The increase in the magnetic field leads to the stiffening of
EOS at lower temperatures, while at higher temperatures, any increase in the magnetic field
results in the softening of EOS.
We have also studied the equation of state for magnetized nuclear matter.
It has been shown that the flashing temperature which corresponds
to the maximum temperature of the nuclear matter at
which the system is self-bound, decreases as the magnetic field
grows. However, our results confirm that the flashing density is not sensitive
to the magnetic field.
The equation of state for magnetized nuclear matter also shows that
a first order liquid gas
phase transition takes place in the nuclear matter.
We have calculated the coexistence curves for the liquid gas
phase transition at different magnetic fields.
We have found that the coexistence
region reduces as the magnetic field grows.
The critical properties of magnetized nuclear matter
at different magnetic fields have also been calculated.
The results indicate that the critical temperature and pressure decrease
by increasing the magnetic field.
Finally, we have presented the order parameter of the liquid gas
phase transition. It was shown that
this quantity decreases as the magnetic field increases.
Our results for the symmetric nuclear matter have important consequences on hot dense magnetized nuclear systems in protoneutron stars and relativistic heavy ion collisions.
%

\acknowledgements{This work has been supported financially by the Center for Excellence in Astronomy and Astrophysics (CEAA-RIAAM).
We also wish to thank the Shiraz University Research Council.}


\begin{table*}
\caption{The flashing temperature, $T_f$, and density, $\rho_f$, for magnetized nuclear matter at different magnetic fields.}
\label{table1}       
\begin{center}  {\footnotesize
\begin{tabular}{|c|c|c|}
\hline{$B \ (G)$} & $T_f (MeV)$  & \multicolumn{1}{c|}{$\rho_f (fm^{-3})$}  \\\hline
$10^{18}$ & 17.5& 0.19 \\
$5\times10^{18}$ &17.4 &0.19   \\
$10^{19}$& 17.0& 0.19 \\

\hline
\end{tabular} }
\end{center}
\vspace*{2cm}  
\end{table*}
\newpage

\begin{table*}
\caption{The critical temperature, $T_c$, density, $\rho_c$, and pressure, $p_c$, for magnetized nuclear matter  at different magnetic fields.}
\label{table2}       
\begin{center}  {\footnotesize
\begin{tabular}{|c|c|c|c|}
\hline{$B \ (G)$} &
\multicolumn{1}{c|}{$T_c (MeV)$}&\multicolumn{1}{c|}{$\rho_c(fm^{-3})$} &\multicolumn{1}{c|}{$p_c (MeV fm^{-3})$} \\\hline
$10^{18}$ & 22.9 &0.14 &0.93      \\
$5\times10^{18}$  &22.6 &0.14  &0.90  \\
$10^{19}$& 21.8& 0.14&0.84 \\
Experimental result  \cite{Elliott}  &17.9 &0.06  &0.31  \\

\hline
\end{tabular} }
\end{center}
\vspace*{2cm}  
\end{table*}

\newpage
\newpage
\newpage
\newpage
\newpage
\newpage

\newpage

\newpage

\newpage

\newpage
\newpage

\newpage

\newpage



%


\end{document}